
\input harvmac


\noblackbox
\baselineskip=16pt

\Title{\vbox{\baselineskip12pt{\hbox{CTP-TAMU-52/92}}%
{\hbox{hep-ph@xxx/9207202}}}}
{\vbox{\centerline{Dilatonic Domain Walls}}}

\centerline{HoSeong ~La\footnote{$^*$}{%
e-mail address: hsla@phys.tamu.edu, hsla@tamphys.bitnet}   }

\bigskip\centerline{Center for Theoretical Physics}
\centerline{Texas A\&M University}
\centerline{College Station, TX 77843-4242, USA}
\vskip 0.7in

A static domain wall with infinitesimal thickness is obtained
in the theory of a scalar field coupled to gravity with
the dilaton, i.e. the Jordan-Brans-Dicke gravity.
The value of the dilaton is determined in terms of the Brans-Dicke
parameter $\omega$.
In particular the metric for the wall becomes flat and the
dilaton, though nontrivial, has vanishing kinetic energy (i.e. $\omega=0$)
 so that there is no gravitational effect due to
such a dilatonic domain wall.
Some remarks on possible thick domain walls are given too.
\baselineskip=16pt     

\bigskip
{\noindent PACS numbers: 04.60.+n, 04.50.+h, 98.80.-k, 11.10.-z,
11.30.Er, 11.17.+y }

\Date{06/92} 
 \noblackbox

\def\tilde{\widetilde}

\def\la{\lambda}
\def\half{{\textstyle{1\over 2}}}

\def\e{{\rm e}}
\def\pa{\partial}
\def\mbox#1#2{\vcenter{\hrule \hbox{\vrule height#2in
		\kern#1in \vrule} \hrule}}  

\def\eps{\epsilon}

\font\cmss=cmss10 \font\cmsss=cmss10 scaled 833
\def\IZ{\relax\ifmmode\mathchoice
{\hbox{\cmss Z\kern-.4em Z}}{\hbox{\cmss Z\kern-.4em Z}}
{\lower.9pt\hbox{\cmsss Z\kern-.4em Z}}
{\lower1.2pt\hbox{\cmsss Z\kern-.4em Z}}\else{\cmss Z\kern-.4em Z}\fi}

\def\CS{{\cal S}}

\vfill\eject


It is interesting to observe the fact that most of gravitational theories
known so far allow extended objects as their classical
solutions\ref\Vilrev{For reviews, see A. Vilenkin, Phys. Rep. {\bf 121}
(1985) 263.}%
\ref\Fbrev{For recent reviews in string theory case, see
M.J.~Duff and J.X. Lu, Class. Quant. Grav. {\bf 9} (1992) 1;
C.G. Callan, J.A.~Harvey and A.~Strominger ``Supersymmetric String Solitons ,''
Chicago preprint, EFI-91-66 (1991).}, although the chances of actual
existence are not necessarily high. Nevertheless, the properties of such
extended objects can often provide constraints on various aspects of
phenomenological implications of the theories. Theories
which undergo phase transition via spontaneous symmetry breaking
are particularly interesting because they
become building blocks of cosmological models and often
provide topological defects, e.g. monopoles, cosmic strings, domain walls,
etc.\Vilrev.  Such defects in principle
can be formed in the early universe\ref\KoTu{E.W. Kolb and M.S.
Turner, {\it`` The Early Universe,"} (Addison-Wesley, New York, 1989).}.

In this letter we shall investigate the structure of (infinitesimally thin)
domain wall
solutions in the Jordan-Brans-Dicke (JBD) theory, which is a gravitational
theory with the dilaton in the four-dimensional space-time.
{}From the string theory's point of view, the JBD theory with a specific
Brans-Dicke (BD) parameter is a natural
effective gravitational theory before the dilaton freezes up.
Furthermore, a cosmological  model can also be built based on the JBD
theory\ref\DLaSt{D. La  and P. Steinhardt, Phys. Rev. Lett. {\bf 62} (1989)
376.}. Thus it is
worth while to investigate the existence of extended objects in this
context. Cosmic string
solutions in this theory were previously studied in ref.\ref\GuOr{C.
Gundlach and M.E. Ortiz, Phys. Rev. {\bf D42} (1990) 2521.}.

We shall consider the action for the JBD theory
\eqn\ei{\CS=\int d^4x\sqrt{-\tilde g}\e^{-2\tilde\phi}\left(\tilde{R}
-4\omega\pa_\mu\tilde{\phi}\pa^\mu\tilde{\phi}\right)
+\CS_M[\tilde{g}_{\mu\nu}],}
where $\omega$ is the BD parameter of the theory.
In particular the $\omega=-1$ case corresponds to the action of the dilaton
gravity from string theory\ref\rdilgr{J. Scherk and J.H. Schwarz, Nucl.
Phys. {\bf B81} (1974) 118\semi E.S. Fradkin and A.A. Tseytlin, Nucl. Phys.
{\bf B261} (1985) 1\semi C.G. Callan , D. Friedan, E. Martinec and M.J.
Perry, Nucl. Phys. {\bf B262} (1985) 593.}.
For our purpose we take $\CS_M$ as the action for the real scalar field
with a double well potential:
\eqn\eii{\CS_M=-\half\int d^4x\sqrt{-\tilde g}\left(\pa_\mu\Phi\pa^\mu\Phi
+\la(\Phi^2-\eta^2)^2\right).}
$\CS_M$ in particular has a discrete symmetry $\Phi\to -\Phi$. In the
static case, this discrete symmetry is equivalent to the time reversal
symmetry. This is why the domain wall structure is related to the CP
phases\ref\TDL{T.D. Lee, Phys. Rep. {\bf 9} (1974) 143.}\ref\rZel{Ya.B.
Zel'dovich, I.Yu. Kobzarev, and L.B. Okun, JETP {\bf 40} (1975) 1.}.
For convenience we redefine the variables as
\eqn\eiii{\omega={1\over 4\beta^2}-{3\over 2},\ \
\e^{-2\tilde\phi}={1\over 16\pi G}\e^{-2\beta\phi},\ \
\tilde{g}_{\mu\nu}=\e^{2\beta\phi} g_{\mu\nu},}
then the action now becomes
\eqn\esa{\CS={1\over 16\pi G}\int d^4x\sqrt{-g}(R-\pa_\mu\phi\pa^\mu\phi)
+\CS_M[\e^{2\beta\phi}g_{\mu\nu}].}
Setting $\phi=0$, this action reduces to that of a real scalar field
coupled to the Einstein gravity, in which  case a static thin
domain wall solution is
known to exist, although it is gravitationally unstable%
\ref\doref{A. Vilenkin, Phys. Rev. {\bf D23} (1981) 852.}.
In this letter we shall present a stable domain wall, which is a static
classical solution of eq.\ei.
In our case although $\tilde{g}_{\mu\nu}$ is a physical metric of the
space-time, eq.\esa\ is very convenient for practical purposes.
Although $\tilde\phi$ is the actual dilaton, we will call $\phi$ a dilaton too
since they are related by a simple field redefinition.

\def\sqg{\sqrt{-g}}
\def\mn{\mu\nu}
\def\tph{\tilde{\phi}}

By varying eq.\esa\ with respect to the new metric $g_{\mu\nu}$, we obtain
\eqn\emet{R_{\mu\nu}=\pa_\mu\phi\pa_\nu\phi+8\pi G(T_{\mu\nu}-\half
g_{\mu\nu}T),}
where the ``energy-momentum'' tensor is given by
\eqn\eten{T_{\mu\nu}=-{2\over\sqrt{-g}}{\delta\CS_M\over\delta g^{\mu\nu}}
=-\half\left[\e^{2\beta\phi}g_{\mu\nu}g^{\alpha\beta}\pa_\alpha\Phi\pa_\beta\Phi
-2\e^{2\beta\phi}\pa_\mu\Phi\pa_\nu\Phi+g_{\mu\nu}\e^{4\beta\phi}\la
(\Phi^2-\eta^2)^2\right].}
Note that $T_{\mu\nu}$ is not conserved due to the dilatonic contribution.
The physical energy-momentum tensor $\tilde{T}^{{\rm
matter}}_{\mn}+\tilde{T}^{\tilde{\phi}}_{\mn}$
satisfies the gravitational equation of motion for JBD theory\ref\Wein{S.
Weinberg, ``{\it Gravitation and Cosmology}," (Wiley, New York, 1972).},
\eqn\jbeom{\tilde{R}_{\mn}-\half\tilde{g}_{\mn}\tilde{R}
=8\pi\e^{2\tilde{\phi}} \left(\tilde{T}^{{\rm
matter}}_{\mn}+\tilde{T}^{\tilde{\phi}}_{\mn}\right),}
where $\nabla^\mu\tilde{T}^{{\rm matter}}_{\mn}=0$ and
\eqn\dilem{\eqalign{
\tilde{T}^{\tilde{\phi}}_{\mn}=&
{1\over 8\pi}\e^{-2\tph}\left(2\pa_\mu\pa_\nu\tph
+2\tilde{\Gamma}^{\alpha}_{\mn}\pa_\alpha\tph+\half\e^{2\tph}\tilde{g}_{\mn}
\tilde{\mbox{.1}{.1}}^2\e^{-2\tph}\right)\cr
&-{\omega\over 2\pi}\e^{-2\tph}
\left(\pa_\mu\tph\pa_\nu\tph-\half\tilde{g}_{\mn}\pa^\alpha\tph\pa_\alpha\tph
\right) .\cr}}
The second term of eq.\dilem\ is proportional to $\omega$ so that it
vanishes if $\omega=0$, but the first term is independent of $\omega$.

The field equations for the dilaton $\phi$ is
\eqn\edil{{\mbox{.1}{.1}}^2\phi={1\over\sqg}\pa_\mu(\sqg g^{\mn}\pa_\nu)\phi=
-8\pi G\beta T}
and the scalar field satisfies
\eqn\ehig{\pa_\mu(\sqg \e^{2\beta\phi} g^{\mn}\pa_\nu)\Phi
-2\la\sqg\e^{4\beta\phi}\Phi(\Phi^2-\eta^2)=0.}

In general, domain wall solutions are obtained in theories where a discrete
symmetry is spontaneously broken. Note that the action eq.\eii\ for the
matter field $\Phi$ has a discrete symmetry $\Phi\to -\Phi$ so that we can
look for domain walls, when this symmetry is spontaneously broken.
In the case where domain walls have infinitesimal
thickness, we can approximate the wanted scalar field to behave as
$$\Phi(z)=\cases{\eta & if $z>0$;\cr -\eta& if $z<0$.\cr}$$
Then we are interested in the the domain wall which
 separates a space of the $\Phi=\eta$ phase from a space of
the $\Phi=-\eta$ phase.
Such an approximation is  in fact reasonable for the cases where the Compton
wavelength of the test particle is much longer than the thickness of the
wall.

We shall try the following {\it ansatz}
for domain wall solutions\ref\rTa{A.H. Taub, Phys. Rev. {\bf 103} (1956) 454.}
\doref:
\eqn\doans{ds^2=A(|z|)(-dt^2+dz^2) +B(|z|)(dx^2+dy^2).}
Note that we have required the reflection symmetry between each side of
 the wall, which is an
infinite plane perpendicular to the $z$-direction at $z=0$.
In fact it turns out that this is quite a general ansatz.
Normally $|z|$ is not analytic at $z=0$, but we can use the following
prescription for $|z|$ to avoid such a difficulty:
\eqn\ezpre{|z|=z[\theta(z)-\theta(-z)],}
  where
$\theta(z)$ is a step function defined by $\theta(z)=1$ for $z\geq 0$,
$\theta(z)=0$ for $z<0$. Then
$\pa_z|z|=[\theta(z)-\theta(-z)]+2z\delta(z)$.
If we promise that $\pa_z|z|$ shall be multiplied
with some function of $z$ that does not have a pole at $z=0$, we
can safely use an identification $\pa_z|z|\equiv\theta(z)-\theta(-z)$.
Similarly,
$\pa_z^2|z|\equiv 2\delta(z)$. The reason we try to be careful about such
analyticity is to check the consistency of the solutions at the wall, which
turns out to be important to
provide interesting constraints on the solutions.

\def\comarg{(1+\tilde{\kappa}|z|)}
Using this ansatz the field equations become
\eqna\comfi
$$\eqalignno{
\left({A'\over A}\right)'+{A'B'\over AB} &=-8\pi G\la
A\e^{4\beta\phi}\left(\Phi^2-\eta^2\right)^2, &\comfi a\cr
\left({B'\over B}\right)'+{B'^2\over B^2} &=-8\pi G\la
A\e^{4\beta\phi}\left(\Phi^2-\eta^2\right)^2, &\comfi b\cr
{A'B'\over AB}\!-\!\!\left({A'\over A}\right)'\!\! \!
-\!2{B''\over B}\!+\!\!{B'^2\over B^2}\!\!
&=\!2\phi'^2\!\!+\!8\pi G\!\left(
2\e^{2\beta\phi}\Phi'^2\!+\!\la A\e^{4\beta\phi}
(\Phi^2\!\!-\!\eta^2)^2\right)\!\!,
&\comfi c\cr
{B'\over B}\phi'+\phi''&=8\pi G\beta\left(\e^{2\beta\phi}\Phi'^2 +2\la
A\e^{4\beta\phi}(\Phi^2-\eta^2)^2\right)\!\!, &\comfi d\cr
0&=\left(B\e^{2\beta\phi}\Phi'\right)' -2\la AB
\e^{4\beta\phi}\Phi(\Phi^2-\eta^2), &\comfi e\cr}$$

For thin walls we have $\Phi^2=\eta^2$ and $\Phi'=0$ away from $z=0$ so
that we shall first solve the above equations away from $z=0$, then shall check
the consistency at the wall.
Solving eqs.\comfi{a\hbox{--}d}\ for $z\neq 0$, we obtain
\eqn\solone{\eqalign{A(z)&=\comarg^{\alpha^2-{1\over 2}},\cr
	B(z)&=\comarg,\cr
	\phi(z)&=\alpha\ln\comarg,\cr}}
where $\tilde\kappa$ and
the parameter $\alpha$ are
yet to be determined, while the value of the dilaton at $z=0$ is taken to be
zero.

Now let us check the consistency of the solution at the wall.
Using the analytic property eq.\ezpre\ we prescribed, eq.\comfi{b} reduces
to
\eqn\corchi{\tilde\kappa 2\delta(z)= 8\pi
G\la\left(\Phi^2-\eta^2\right)^2, }
which leads to $\tilde\kappa>0$. Similarly,  eq.\comfi{a}\ reduces
to
\eqn\conchi{\left(\alpha^2-\half\right)\tilde\kappa 2\delta(z)= 8\pi
G\la\left(\Phi^2-\eta^2\right)^2,}
which together with eq.\corchi\ implies that
$$\alpha^2=3/2\ \ {\rm  so\ \  that}\ \ A=B.$$
This is a very strong constraint, which even the Einstein gravity case
cannot satisfy\foot{$\alpha=0$ corresponds to the Einstein gravity case.
This also confirms that  there is no static (stable) thin domain wall in this
case. Note that this does not imply that there is no domain wall solution in
the Einstein gravity. It is known that there exits a time-dependent thin wall
solution: A.~Vilenkin, Phys. Lett. {\bf 133B} (1983) 177; J.~Ipser and
P.~Sikivie, Phys. Rev. {\bf D30} (1984)712.}.
Note that $\tilde\kappa$ and $\eta$ have mass dimensions and the
Newton's gravitational constant $G$ has inverse mass square dimension, while
$\la$ is a dimensionless coupling constant. Using a dimensional analysis for
a possible thick wall, if we have
$G\la\eta^4\gg \tilde{\kappa}{\eta\over\sqrt\la}$ and $\Phi(z=0)=0$,
the LHS of eqs.\corchi\conchi\ are effectively
comparable to the RHS by smearing out the delta function
Thus this is a good approximate solution and
the condition in fact corresponds to the weak gravitational field limit.

Eqs.\comfi{c,d}\ in turn require that
\eqn\eI{(\Phi')^2={\tilde\kappa\over 4\pi G}\delta(z)}
and
\eqn\eII{2\beta\alpha^2+\alpha=0.}
Thus $2\alpha\beta+1=0$ and that eq.\comfi{e}\ is also satisfied.
For $\alpha^2=3/2$, $\beta^2=1/6$, which leads to the BD parameter
$\omega=0$ for eq.\eiii.

Finally, $\tilde\kappa$ can be determined from the ``energy''
density as follows:
Using eqs.\corchi -\eII\ we can compute the ``energy-momentum'' tensor
eq.\eten\
as
\eqn\edisen{T_{\mu\nu}={\tilde\kappa\over 4\pi G}\delta (z)
{\rm diag} (1, -1, -1, 0),\ \ \ \ \tilde\kappa>0.}
Note that this is not the physical energy-momentum in eq.\jbeom\ so that
getting negative ``energy'' density does not imply the instability of the wall.

Here we would like to call the readers attention to the fact that we have
differential equations with the Dirac delta-function. Some may find that this
is unreasonable because after all the Dirac delta-function is not a function
but a distribution. But this is not completely unreasonable in field theory
when we often need to be careful about the analyticity. The main intention is
not to solve the differential equations in question
but to check the consistency between
equations. In this sense this is a sufficiently good approximation.
In fact one can be more careful about this situation and can introduce
distributional energy-momentum tensor in terms of delta-function from the
beginning\ref\rIsGer{W. Israel, Nuo. Cim. {\bf 44B} (1966) 1;
R. Geroch and J. Traschen, Phys. Rev. {\bf D36} (1987) 1017.}.
The result however is more or less equivalent because we also have
derived the distributional ``energy-momentum'' tensor using our prescription
eq.\ezpre. We can also further clarify the
result by introducing infinitesimal thickness of the wall and taking
approximation around $z=0$, although we cannot determine the shape of the
solution within this thickness exactly%
\foot{For example, we can cut off the solution
for $|z|>\eps$ for infinitesimal $\eps$. Although we
are not able to solve the equations
exactly in the region $|z|<\eps$ but at least we know $\Phi(0)=0$ and $A, B,
\phi$ should be continuous. Then we can check the consistency of the solution
eq.\solone\ at $z=0$ because as $\eps\to 0$ the leading
order of the solution should not be much different from this. In fact we only
need to use the property $A=B^{\alpha^2-1/2}$ and $\phi=\alpha\ln B$.
Thus it is a good approximation.}.

Now the physical metric can be obtained by multiplying the conformal factor
(see eq.\eiii) as $d\tilde{s}^2=\e^{2\beta\phi}ds^2$ so that we obtain
\eqn\resone{d\tilde{s}^2=\comarg^{2\alpha\beta+\alpha^2-{1\over 2}}(-dt^2+dz^2)
+\comarg^{2\alpha\beta+1}(dx^2+dy^2).}
Since $2\alpha\beta+1=0$, we have $\omega=0$, then the physical metric
becomes nothing but the Minkowski metric
\eqn\restwo{d\tilde{s}^2=-dt^2+dz^2+dx^2+dy^2.}


The success of the prescription eq.\ezpre\ indeed suggests that we should be
able to get thick wall solutions by replacing the step function $\theta (z)$
with a smeared $\theta$-function\ref\rLaprom{H.S. La, to be published.}
Thick wall solutions we are really interested in nevertheless should
approach to our thin wall solution asymptotically as $|z|\to\infty$.
Also in the the small $\beta$ approximation,
if $\beta\to 0$, $\omega\to \infty$, and that
$\phi\to 0$, it should recover a thick domain wall in
the Einstein gravity case. But solving eq.\comfi{a-e}\
for an arbitrary $\beta$
and a wall thickness seems to be fairly complicated, though it may not be
impossible.

\bigskip
\leftline{\bf Discussion}
\medskip

Needless to say such a thin dilatonic domain wall is gravitationally stable
because it does not carry any gravitational energy-momentum (see
eq.\dilem). The dilatonic contribution
to the energy-momentum tensor precisely cancels the matter part of
the energy-momentum tensor to make the space-time flat.
It should be important to further look for time-dependent or thick wall
solutions to check if such a result survives in these cases too.

This result can be very significant in the following sense:

First, such a dilatonic domain wall may not be seriously
harmful with respect to the strong CP problem.
The domain wall problem in the strong CP problem is generated by the
existence of a domain wall that carries too much energy. But the dilatonic
domain wall we have at hand does not carry any gravitational energy so that
it can exist without ruining the standard cosmology paradigm.

Second, it has been speculated that the extended object in string theory,
in particular, the fivebranes%
\ref\Duf{%
M.J. Duff, Class. Quan. Grav. {\bf 5} (1988) 189\semi
M.J. Duff, in {\it Superworld II}, ed. by A. Zichichi (Plenum, New York, 1990).
}\ref\Stro{A. Strominger, Nucl. Phys. {\bf B343} (1990) 167.},
can appear as a two-dimensional extended objects in the four-dimensional
space-time after compactifications%
\ref\Lasolb{H.S. La, ``Solitons reduced
from Heterotic Fivebranes," CTP-TAMU-36/92 (1992).}.
The metric given in ref.\Lasolb\ for the sine-Gordon fivebrane is in fact
flat because the curvature vanishes.
 Since the dilaton gravity is a natural gravitational theory from the
string theory's point of view, the domain wall for this theory, if it ever
exists, may be in such a form. Domain walls in string theory context were
also
formerly studied in other aspects\ref\rdomst{K. Choi and J.E. Kim, Phys.
Rev. Lett. {\bf 55} (1985) 2637\semi J.A. Casas and  G.G. Ross, Phys.
Lett. {\bf B198} (1987) 461\semi M. Cvetic, F. Quevedo and S.J. Rey, Phys.
Rev. Lett. {\bf 67} (1991) 1836.}\ so that it would be interesting to
further study with such an expectation presented here.

{}From such a point of view one may be tempted to
speculate that there may not be a domain wall problem
in string theory, although a domain wall exists.
Since our conclusion can be applied to a possible thick domain wall case
far away from the wall, if such a domain  wall exists, say,  near the
horizon of the universe, we certainly cannot observe any gravitational
effects due to the wall.

\bigbreak\bigskip\bigskip\centerline{{\bf Acknowledgements}}\nobreak

\par\vskip.3truein

The author would like to thank A. Vilenkin for very helpful
discussions and C.~Vafa
for his hospitality at Harvard, while this work was done.
This work was supported in part by NSF grant PHY89-07887 and a World Laboratory
Fellowship.


%
\listrefs
\vfill\eject
\bye